# Effect of magnetic (Nd) doping on electrical and magnetic properties of topological $Sb_2Te_3$ single crystal

Kapil Kumar[1,2], Yogesh Kumar[1,2], M. Singh[3], S. Patnaik[3], I. Felner[4] and V.P.S. Awana[1,2*]

[1]CSIR-National Physical Laboratory, Dr. K. S. Krishnan Marg, New Delhi-110012, India
[2]Academy of Scientific and Innovative Research (AcSIR), Ghaziabad 201002, India
[3]School of Physical Sciences, Jawaharlal Nehru University, New Delhi, 110067, India
[4]Racah Institute of Physics, The Hebrew University, 91904, Jerusalem, Israel

**Abstract**

Here, we report the growth and characterization of single crystals of $Nd_xSb_{2-x}Te_3$ (x = 0 & 0.1), by solid state reaction route via self-flux method. The phase and layered growth are confirmed through X-ray diffraction and Scanning electron microscopy respectively. A slight contraction in lattice parameters is seen after Nd doping. Also a minute shift in vibrational modes of recorded Raman spectra has been observed by doping of Nd in $Sb_2Te_3$. The magneto-resistance values under magnetic field of 5Tesla for $Sb_2Te_3$ are 75% at 2.5K and 60% at 20K, but only 40% at 5K for $Nd_{0.1}Sb_{1.9}Te_3$. DC magnetic measurements exhibit expected diamagnetic and paramagnetic behaviors for pure and Nd doped crystals respectively. A cusp-like behavior is observed in magneto-conductivity of both pure and Nd doped crystals at low magnetic fields (< 1 Tesla) which is analyzed using Hikami-Larkin-Nagaoka (HLN) model. For $Sb_2Te_3$ the fitted parameters α are -1.02 and -0.58 and the phase coherence lengths $L_\varphi$ are 50.8(6)nm & 34.9(8)nm at temperatures 2.5K and 20K respectively. For $Nd_{0.1}Sb_{1.9}Te_3$, α is -0.29 and $L_\varphi$ is 27.2(1)nm at 5K. The α values clearly show the presence of weak anti-localization effect in both, pure and Nd doped samples. Also with Nd doping, the contribution of bulk states increases in addition to conducting surface states in overall conduction mechanism.

[*]**Corresponding Author**
Dr. V. P. S. Awana: E-mail: awana@nplindia.org
Ph. +91-11-45609357, Fax-+91-11-45609310
Homepage: awanavps.webs.com

**Introduction:**

Topological insulators (TI) are the quantum materials possessing insulating bulk states and conducting surface states (SS), which results from band inversion due to presence of strong spin-orbit coupling in the material [1-3]. These surface states exhibit spin-momentum locking and are topologically protected by time reversal symmetry which leads to absence of backscattering from non-magnetic impurities [4-6]. TIs show unique quantum properties such as: observation of high magneto-resistance, quantum spin Hall effect, and terahertz generation etc. It is also reported that in topological superconductors, the proximity effect can lead to the observation of Majorana fermions [7-10]. However, doping of TIs with magnetic atoms will leads to interesting phenomena as a consequence of the interaction between topology and magnetism in such materials. The spontaneous magnetization of magnetic dopants breaks the time reversal symmetry, which leads to the opening of energy gap at Dirac point [11]. These materials are predicted to host peculiar properties such as topological magneto-electric effect, image magnetic monopoles, Majorana fermions, and quantum anomalous Hall effect (QAHE) etc [12-15]. The QAHE, first theoretically and then experimentally observed in magnetic doped TIs, can be used for fabrication of dissipation-less electronic devices, spintronics or in quantum computing [16]. Recent studies on doping of bismuth-based chalcogenides $Bi_2Te_3$ and $Bi_2Se_3$, by $3d$ transition elements reported the observation of QAHE in these materials at milli-Kelvin temperatures [17-19].

However, for practical applications it is required to observe QAHE at high temperatures and with reduced concentration of magnetic doping. These conditions can be reached by doping TIs with high magnetic moment elements. Previous reports suggested that the opening of energy gap at Dirac points due to magnetic doping is proportional to the magnetic moment of dopant element [20]. In this context, doping of TIs with transition metal ions can be replaced by rare earth elements. The intrinsic large magnetic moment of rare earth dopants can be useful in two ways: first it can induce further more energy gap at Dirac point and secondly, the doping concentration can be decreased [21-23]. In TIs, the evidence of topological SS and the effect of magnetic doping on electronic properties such as band opening at Dirac point have been studied by angle resolved photoemission spectroscopy [23, 24]. The π Berry phase acquired by an adiabatically moving electron around Fermi surface leads to the destructive interference between two time-reversed scattering paths. This destructive interference enhances the conductivity with decreasing temperature, which is the signature of weak anti-localization (WAL) effect [25, 26]. The magneto-conductivity of TIs has been studied by using Hikami-Larkin-Nagaoka (HLN) model. The fitting parameter α and $L_\varphi$ gives the information about presence of WAL or weak localization and phase coherence length [27, 28].

Based on first principle electronic structure calculations, $Sb_2Te_3$ is a known TI possessing topological SSs. which are protected by TRS. The evidence of these SSs and nontrivial topology in $Sb_2Te_3$ has already been confirmed experimentally from detailed Angle-resolved photoemission spectroscopy (ARPES) exhibiting that $Sb_2Te_3$ belongs to Z2 TI consisting of single Dirac cone around the Fermi surface [29-31]. In this article we report the growth and characterization of $Nd_xSb_{2-x}Te_3$ topological single crystals with doping concentration x = 0 & 0.1. The structure and the absence of any additional impurity in the grown single crystals have been confirmed using X-ray diffraction (XRD) spectra. The layered growth was observed through Field emission scanning electron microscopy (FESEM) images. Raman spectra are recorded at room temperature to study the different vibrational modes. DC magnetic measurements confirm the paramagnetic (PM) behavior for the Nd doped crystal. The magneto-resistance was measured in a field range of ±5Tesla at 2.5K and 20K for x = 0 and at 5K for x = 0.1. The magneto-conductivity data has been fitted by using HLN model and the fitting parameters α and phase coherence length $L_φ$ were extracted. The observed cusp in magneto-conductivity at low field indicates the presence of weak anti-localization effect.

**Experimental details:**

Single crystals of $Nd_xSb_{2-x}Te_3$ with concentration x = 0 and x = 0.1 were synthesized by self-flux method via solid state reaction route. Stoichiometry ratio of high purity (99.999%) metals of neodymium (Nd), antimony (Sb) and tellurium (Te) powders were taken and ground thoroughly in MBRAUN glove box in argon atmosphere. The grounded powders were then palletized with hydraulic pressure palletizer at a pressure of 50 gm/cm$^3$ and the pellets were vacuum sealed in a quartz ampoule under a pressure of $5×10^{-5}$Torr. The vacuum encapsulated ampoules were placed into PID controlled muffle furnace, heated to 850ºC at a rate of 60ºC/h and then kept for 48 hours in order to get homogeneous molten mixtures. For growth of single crystals, the sample then slowly cooled down to 600ºC at a rate of 2ºC/h and kept there for 24 hours so that the atoms attain their lowest energy positions to meet crystallinity. Finally, the sample was naturally cooled down to room temperature and silvery shiny crystals were obtained. Figure 1 shows the schematic of heat treatment, representing all steps of growth viz. heating, hold at high temperature and slow cooling. The inset shows the image of flat surface of grown crystal (x= 0 and 0.1) being mechanically cleaved along its growth axis. Rigaku made Mini Flex II X-ray diffractometer having Cu-K$_α$ radiation of wavelength 1.5418Å was used to obtain the powder XRD spectra of as grown $Nd_xSb_{2-x}Te_3$ single crystals. FullProf Software was used for Rietveld refinement of the powder XRD data. VESTA software was used to extract unit cell based on refined parameters from Rietveld analysis to visualize atomic structure of the as grown $Nd_xSb_{2-x}Te_3$ single crystals. FESEM was used to visualize the morphology of as grown $Nd_xSb_{2-x}Te_3$ single crystals. Presence of all constituent elements in desired stoichiometric ratio of as grown $Nd_xSb_{2-x}Te_3$ single crystals were confirmed by Energy Dispersive X-ray Analysis (EDAX). Raman spectrums of both crystals were recorded at room temperature for a wavenumber range of 55-300cm$^{-1}$ using Renishaw inVia Reflex Raman Microscope. A 514nm Laser of power less than 5mW was used for measurement to avoid any local heating of sample. All electrical and magnetic measurements of $Nd_xSb_{2-x}Te_3$ single crystals were performed on Quantum Design PPMS (Physical Property Measurement System). The conventional four probe geometry was used for Resistivity vs Temperature measurements on PPMS.

**Results and Discussion**

Powder XRD spectra were recorded on crushed $Nd_xSb_{2-x}Te_3$ single crystals with concentrations x = 0 & 0.1 at room temperature. Figure 2(a) shows the Rietveld refined data of recorded XRD spectra using FullProf software. Red and black curves represent the experimental observed and calculated intensity, whereas blue curve shows the difference between observed and calculated intensity. The Bragg's positions are represented by vertical green bars. The $Nd_xSb_{2-x}Te_3$ belongs to $R\bar{3}m$ space group with rhombohedral crystal structure as suggested by earlier reports [32]. The Rietveld refinement of the lattice parameters are represented in Table 1. Within the uncertainties the lattice parameters remain unchanged. No additional peak of different phases has been observed, which clearly indicates the phase purity of as grown crystals. The crystal structure of $Nd_xSb_{2-x}Te_3$ has obtained using VESTA software as shown in Figure 1(b). It consists of layered structure with Te-Sb-Te-Sb-Te stacking. FESEM was performed to determine the surface morphology of mechanically cleaved flakes of as grown crystals. The FESEM images are shown in Figure 3, the bright and dark contrast indicates the layered growth of both crystals. Figure 4 plots the recorded Raman spectra of $Sb_2Te_3$ and Nd doped $Sb_2Te_3$ single crystals measured at room temperature. The black and red curves represent the obtained vibrational modes for concentration x = 0 and 0.1 respectively (the data was plotted using y-offset). Both crystals shows three different vibrational modes with wavenumber ranges from 55-300cm$^{-1}$ and these modes are identified as $A_{1g}^1$, $E_g^2$ and $A_{1g}^2$. For $Sb_2Te_3$, the $A_{1g}^1$, $E_g^2$ and $A_{1g}^2$ modes are observed at 68.6cm$^{-1}$, 111.5cm$^{-1}$ and 165.7cm$^{-1}$ respectively. Whereas for Nd doped $Sb_2Te_3$, the same modes are observed at 70.3cm$^{-1}$, 113.4cm$^{-1}$ and 167.8cm$^{-1}$ respectively. It is observed that with incorporation of Nd doping, all the vibrational modes of Sb$_2$Te$_3$ has been shifted slightly; which clearly indicates the substitutional doping of Nd atoms at Sb sites in Sb$_2$Te$_3$ crystal structure.

The magnetic properties of Sb$_2$Te$_3$ and Nd doped Sb$_2$Te$_3$ are shown in Figure 5(a) and 5(b) respectively. The magnetic behaviour of Sb$_2$Te$_3$ is well studied in the past [39]. We measured the temperature dependent magnetization M(T) in the field-cooled (FC) mode and for the sake of brevity, we show in fig. 5(a) the low range of data for Sb$_2$Te$_3$ measured at 2.5kOe only. This M(T) curve exhibit a typical PM shape and adhere closely to the Curie-Weiss (CW) law: $\chi(T) = \chi_0 + \frac{C}{T-\theta}$ where, $\chi$(=M/H), $\chi_0$ is the temperature independent part, $C$ is the Curie constant, and $\theta$ is the CW temperature. The PM parameters are: $\chi_0$ =2.3×10$^{-3}$emu/mol Oe, C = 8.9×10$^{-4}$emu K/mol Oe and $\theta$ = 0.360(20)K. This C value corresponds to a PM effective moment of P$_{eff}$ = 0.085μ$_B$/formula unit, a value which is very similar that reported in [39]. Worth noting is the fact that $\theta$ is almost zero, indicating the absence of extra magnetic phases. The isothermal magnetization M(H) curve of Sb$_2$Te$_3$ measured at 5K is shown in the inset of fig 5(a). M first increases sharply up to around 2.5kOe and then decreases and becomes negative. This curve can be fitted as: M(H) = M$_S$ + $\chi_d$H, where M$_S$ is the spontaneous magnetization and $\chi_d$H is the linear diamagnetic intrinsic contribution ($\chi_d$ = -8×10$^{-7}$emu/g) which probably stems from the sample holder. The small M$_S$ = 0.0069(10 emu/g is attributed to a tiny ferromagnetic impurity as an extra phase, not detectable by all other methods used here.

The magnetic properties of Nd$_{0.1}$Sb$_{1.9}$Te$_3$ are displayed in Fig. 5(b). The presence of the PM Nd ions is well demonstrated. M(T) measured under 250Oe up to 300K show a PM behaviour which is well fitted with CW law. The PM parameters extracted are: $\chi_0$ = 3.1×10$^{-6}$ emu/mol Oe, C = 0.019emu K/mol Oe and $\theta$ = -19K. Note the relative high negative value of $\theta$. This C value is two orders of magnitude higher than that of Sb$_2$Te$_3$. On the other hand, the expected C value for x = 0.1 PM Nd$^{3+}$ ions are 0.17emu K/molOe. That means that only 12% of Nd ions contribute to the PM behaviour of Nd$_{0.1}$Sb$_{1.9}$Te$_3$ material, a phenomenon which is well accepted for TIs. In contrast to Figure 5(a) the M(H) curve at 5 and 50K (Fig. 5b inset) clearly shows a typical PM behaviour up to 50kOe. Here, the negative contribution observed in Fig 5(a) is negligible.

The curves of normalized resistivity ρ(T)/ρ(150K) for $Sb_2Te_3$ (Red) and $Nd_{0.1}Sb_{1.9}Te_3$ (Black) are shown in Figure 6. In both materials, resistivity increases with increasing temperature measured up to 150K specifying the metallic behaviour. In $Nd_{0.1}Sb_{1.9}Te_3$, the normalised resistivity at lower temperature showed a slight increment, this may be resulted from the increase in degree of disordering due to addition of rare earth element in Sb2Te3.

For magneto-transport studies of $Sb_2Te_3$ and $Nd_{0.1}Sb_{1.9}Te_3$ single crystals, the percentage magnetoresistance (MR) at each temperature has calculated by using the formula MR% = [(ρ(H)-ρ(0))/ρ(0)]×100, where ρ(0) and ρ(H) are the resistivity at zero and applied magnetic field respectively. Figure 7(a) shows the measured percentage transverse magnetoresistance as function of applied magnetic field varying up to ±5Tesla. The MR of x = 0 is measured at 2.5K & 20K and for x = 0.1, at 5K only. At 2.5K, the MR of $Sb_2Te_3$ increases linearly with H, not-saturated and reaches ≈75% at 5Tesla, whereas, at 20K, MR decreases and reaches 60% at 5Tesla. It is also observed that the sharp dip at low magnetic field became broader as the temperature is increased. MR of $Nd_{0.1}Sb_{1.9}Te_3$ shows similar temperature dependence. At 5K, MR reaches only 40% at 5Tesla, which is much smaller than the MR values of $Sb_2Te_3$. Both $Sb_2Te_3$ and $Nd_{0.1}Sb_{1.9}Te_3$ single crystals show dip-like behavior at low magnetic field and positive magnetoresistance, which suggests the presence of weak anti-localization effect in both materials [33, 34]. Also, $Sb_2Te_3$ has sharper dip-like behavior than that of $Nd_{0.1}Sb_{1.9}Te_3$ crystal, which may indicate contribution of more bulk states in the magneto-transport of $Nd_{0.1}Sb_{1.9}Te_3$ [28].

In order to deduce the parameters that characterize the WAL effect, the magneto-conductivity (MC) has been studied. Figure 7(b) shows the calculated magneto-conductivity of $Nd_xSb_{2-x}Te_3$ single crystals in a field range of ±5Tesla. The red and blue curve corresponds to the MC of $Sb_2Te_3$ at temperature 2.5K and 20K respectively, whereas green curve corresponds to MC of $Nd_{0.1}Sb_{1.9}Te_3$ at 5K. At lower magnetic field, a sharp cusp-like behavior is observed, which broadens with increase in temperature. Both, negative MC and cusp-like behavior corresponds to presence of WAL effect [35]. The Hikami-Larkin-Nagaoka (HLN) model has been used to analyze the WAL effect in topological insulators. According to this model [36], the quantum correction to 2D magneto-conductivity can be explained as

$$\Delta\sigma(H) = -\frac{\alpha e^2}{\pi h}\left[ln\left(\frac{B_\varphi}{H}\right) - \Psi\left(\frac{1}{2} + \frac{B_\varphi}{H}\right)\right]$$

Where, $\Delta\sigma(H)$ is given as difference between magneto-conductivity at applied and zero magnetic field i.e., $\sigma(H) - \sigma(0)$. $\Psi$ is digamma function, h is Plank's constant, e is the electronic charge, H is applied magnetic field, $B_\varphi = \frac{h}{8e\pi L_\varphi^2}$ is characteristic field, α represents number of conduction channel, and $L_\varphi$ is phase coherence length. The extracted pre-factors α and $L_\varphi$ are listed in Table 2. For a system with strong spin-orbit coupling, the pre-factor α will contribute -0.5 for each conduction channel. It also gets values of -0.5 and 1 for weak anti-localization and weak localization respectively [37, 38]. Figure 7(b) (black solid) shows the HLN fitting of magneto-conductivity data in low field regime (-1 to 1T), red and blue symbols represent the obtained MC for $Sb_2Te_3$ and green symbols for $Nd_{0.1}Sb_{1.9}Te_3$ single crystals. The obtained α and $L_\varphi$ values are represented in Table 2. At 2.5K, the pre-factor α = -1.02 is nearly equal to -1, which suggests contribution from two conducting channels in overall magneto-conductivity. With increasing temperature to 20K, the pre-factor α value increased to -0.58 (close to 0.5) which indicates contribution from a single coherent conducting channel and $L_\varphi$ value decreased from 50.8(6)nm to 34.9(8)nm. MC of $Nd_{0.1}Sb_{1.9}Te_3$ shows similar field dependence as that of $Sb_2Te_3$. At 5K, HLN fitted parameters are smaller: α = -0.29 and phase coherence length $L_\varphi$ = 27.2(1)nm. The α value is in between 0 and -0.5, suggests that in addition to surface states there is also a positive contribution from bulk states to the magneto-conductivity. Table 2 shows that all α values obtained are negative, which suggests the presence of weak anti-localization effect in these materials.

**Conclusion**

In summary, we have synthesized pure $Sb_2Te_3$ and $Nd_{0.1}Sb_{1.9}Te_3$ single crystals. In XRD spectra, the absence of any additional peak confirms the phase purity. In Raman spectra, the obtained vibrational modes signify the doping of Nd atoms at Sb sites. Both samples show high positive magnetoresistance at all measured temperatures, signifies the presence of weak anti-localization effect. The magneto-conductivity is analyzed using HLN model. At 2.5K for $Sb_2Te_3$ there is contribution from more than one conduction channel in overall conductivity, which further suppressed to one conduction channel as temperature is increased to 20K. For $Nd_xSb_{2-x}Te_3$ there is contribution from both surface states and bulk states in overall conduction. Magnetization studies indicate clearly, that a small fraction of Nd ions contribute to the PM nature of $Nd_xSb_{2-x}Te_3$, a phenomenon which is typical to TIs materials.

**Acknowledgment**

The authors would like to thank the Director of National Physical Laboratory (NPL), India, for his keen interest in the present work. Also, author like to thank UGC and CSIR, India, for a research fellowship, and AcSIR-NPL for Ph.D. registration.

**Table 1:** Lattice parameters of $Sb_2Te_3$ and $Nd_{0.1}Sb_{1.9}Te_3$ obtained from Rietveld refinement of powder X-ray diffraction spectra.

|  | **a(Å)** | **b(Å)** | **c(Å)** |
|---|---|---|---|
| **$Sb_2Te_3$** | 4.267(2) | 4.267(2) | 30.468(5) |
| **$Nd_{0.1}Sb_{1.9}Te_3$** | 4.261(5) | 4.261(5) | 30.421(1) |

**Table 2:** The pre-factor α and phase coherence length ($L_\varphi$) extracted by HLN fitting of magneto-conductivity of $Sb_2Te_3$ and $Nd_{0.1}Sb_{1.9}Te_3$ up to 1T.

|  | **Temperature (K)** | **α** | **$L_\varphi$ (nm)** |
|---|---|---|---|
| **$Sb_2Te_3$** | 2.5 | -1.02 | 50.8(6) |
|  | 20 | -0.58 | 34.9(8) |
| **$Nd_{0.1}Sb_{1.9}Te_3$** | 5 | -0.29 | 27.2(1) |

**Figure captions:**

**Figure 1:** Solid-state reaction method used by following the shown heat treatment to grow $Nd_xSb_{2-x}Te_3$ crystals for x = 0 and x = 0.1. Inset shows the obtained silvery shining single crystals.

**Figure 2: (a)** Rietveld refined X-ray diffraction spectra for x = 0 and x = 0.1, red and black curve shows the obtained and calculated intensity respectively. Blue curve is the difference between them and green bars represents the Bragg's positions. **(b)** Crystal structure obtained from VESTA software.

**Figure 3:** Field emission scanning electron microscopy images of $Sb_2Te_3$ and $Nd_{0.1}Sb_{1.9}Te_3$ shows the layered growth of obtained single crystals.

**Figure 4:** Raman spectra for x = 0 and 0.1 crystals in a range of 55-300 $cm^{-1}$ measured at room temperature.

**Figure 5:** Temperature dependence of magnetization for moments **(a)** $Sb_2Te_3$ and **(b)** $Nd_{0.1}Sb_{1.9}Te_3$. The insets show the isothermal M(H) curves.

**Figure 6:** Red and blue curve shows the normalized resistivity $\rho(T)/\rho(150K)$ versus temperature plot of $Nd_xSb_{2-x}Te_3$ crystals for x = 0 and x = 0.1 respectively at zero field.

**Figure 7: (a)** Magnetoresistance of pure and Nd doped $Sb_2Te_3$ at different temperatures for a field range of ±5Tesla. **(b)** Red, blue curve represents MC for x = 0 and green curve represents the MC for x = 0.1. Black solid curve shows the low field (up to 1Tesla) HLN fitting of MC at different temperatures.

Fig 1

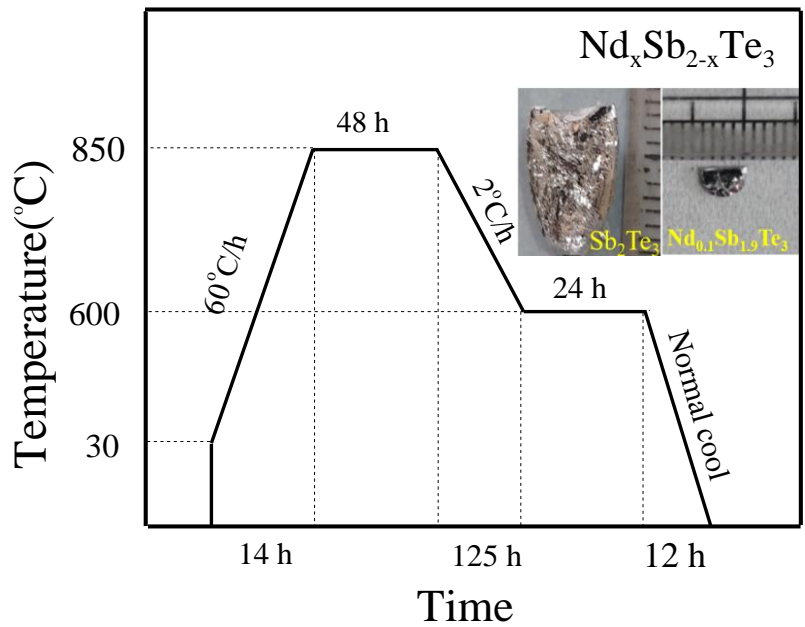

Fig 2(a)

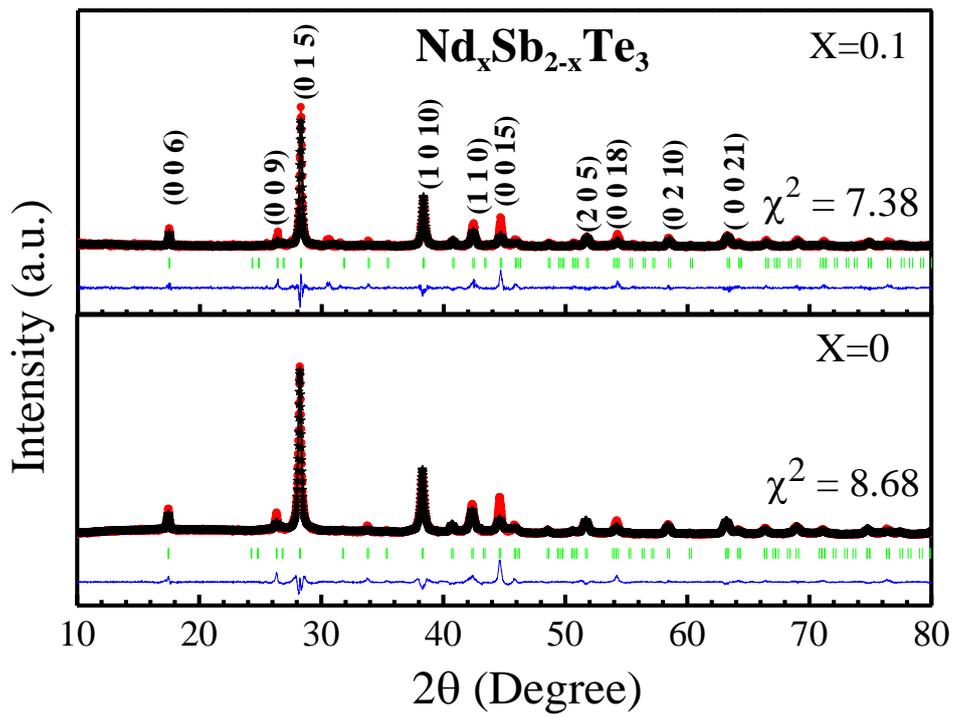

Fig 2(b)

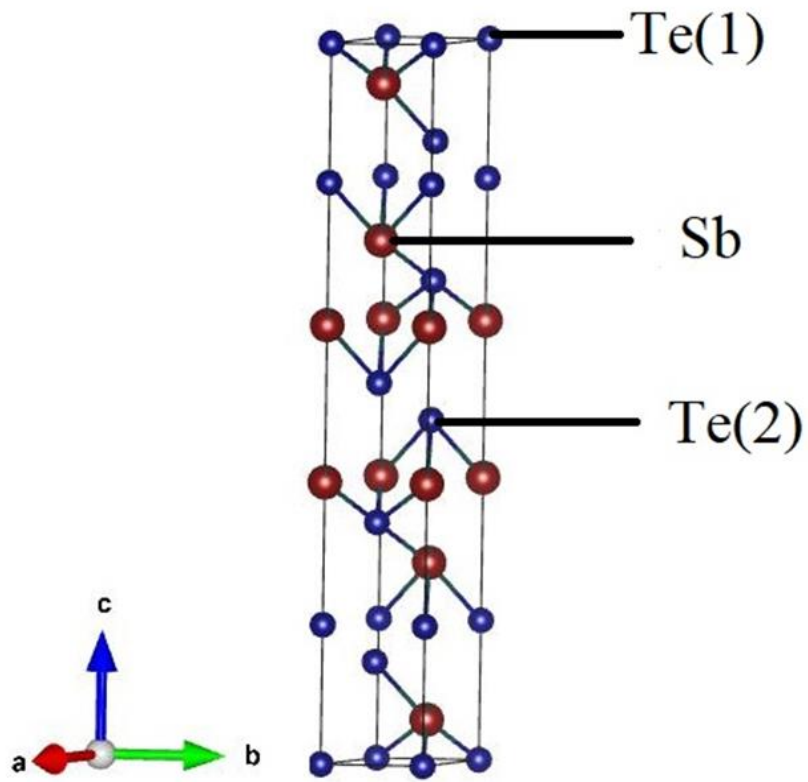

Fig 3

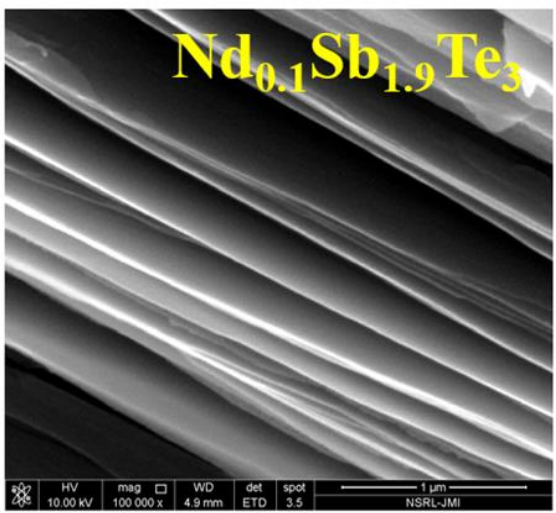 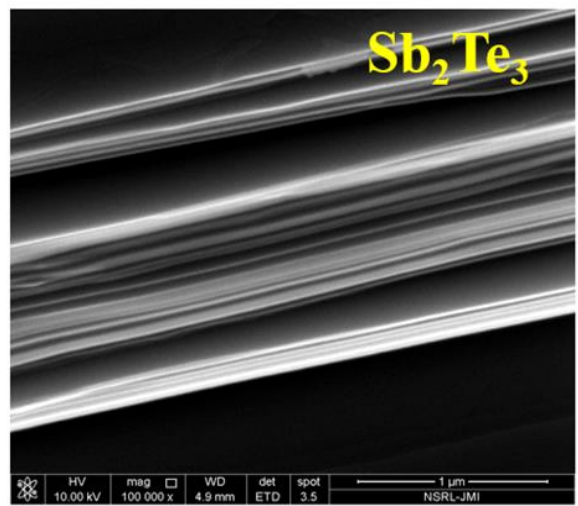

Fig 4

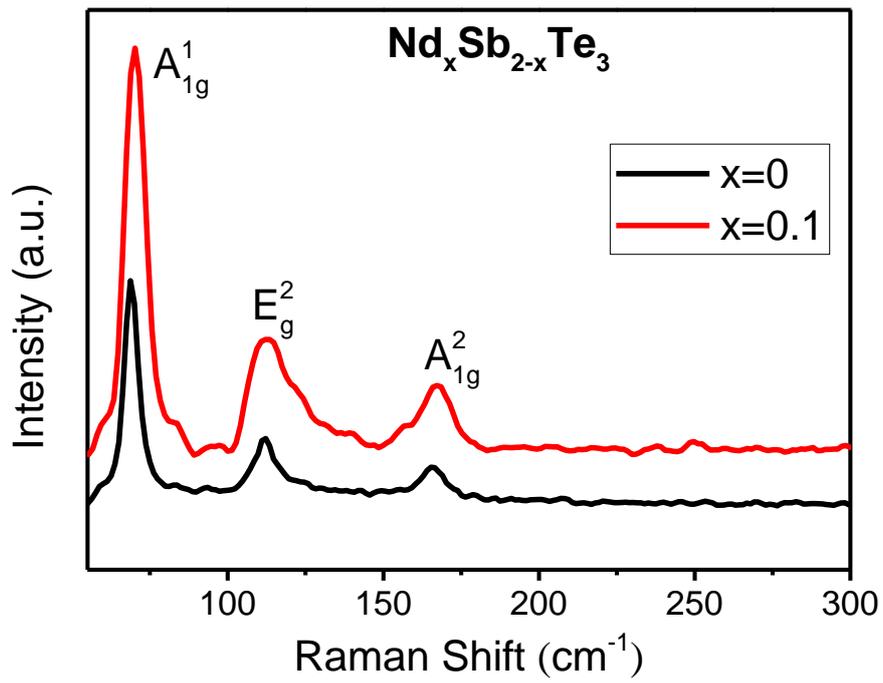

Fig 5(a)

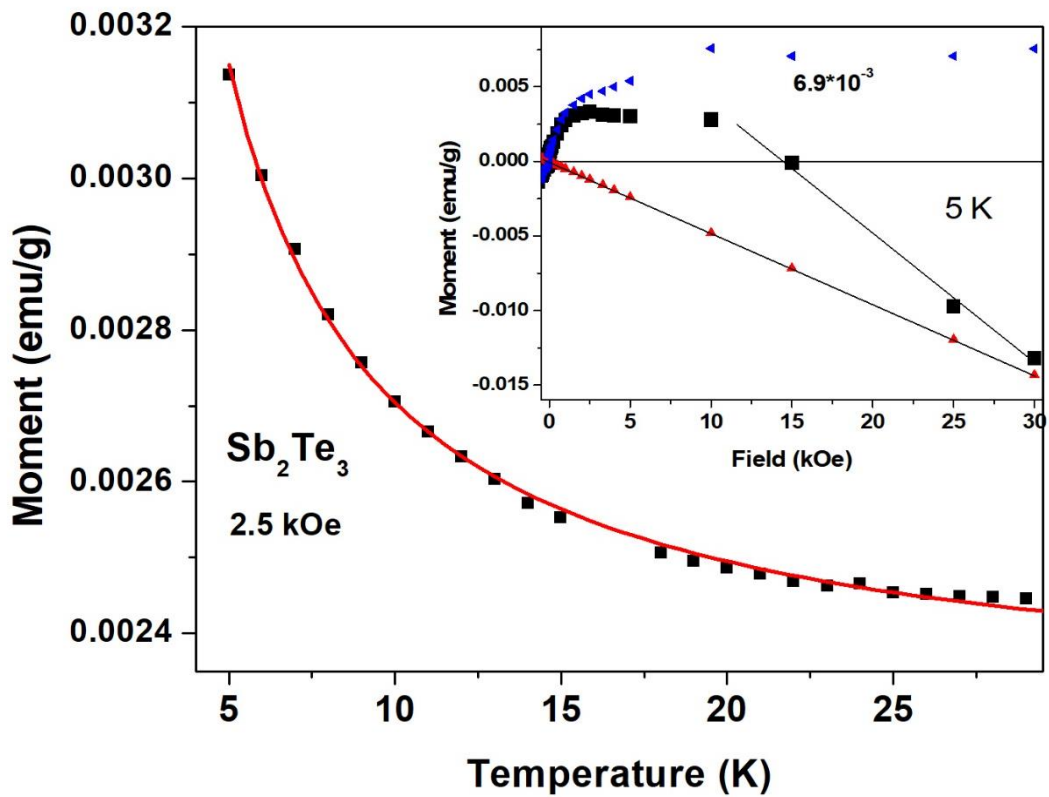

Fig 5(b)

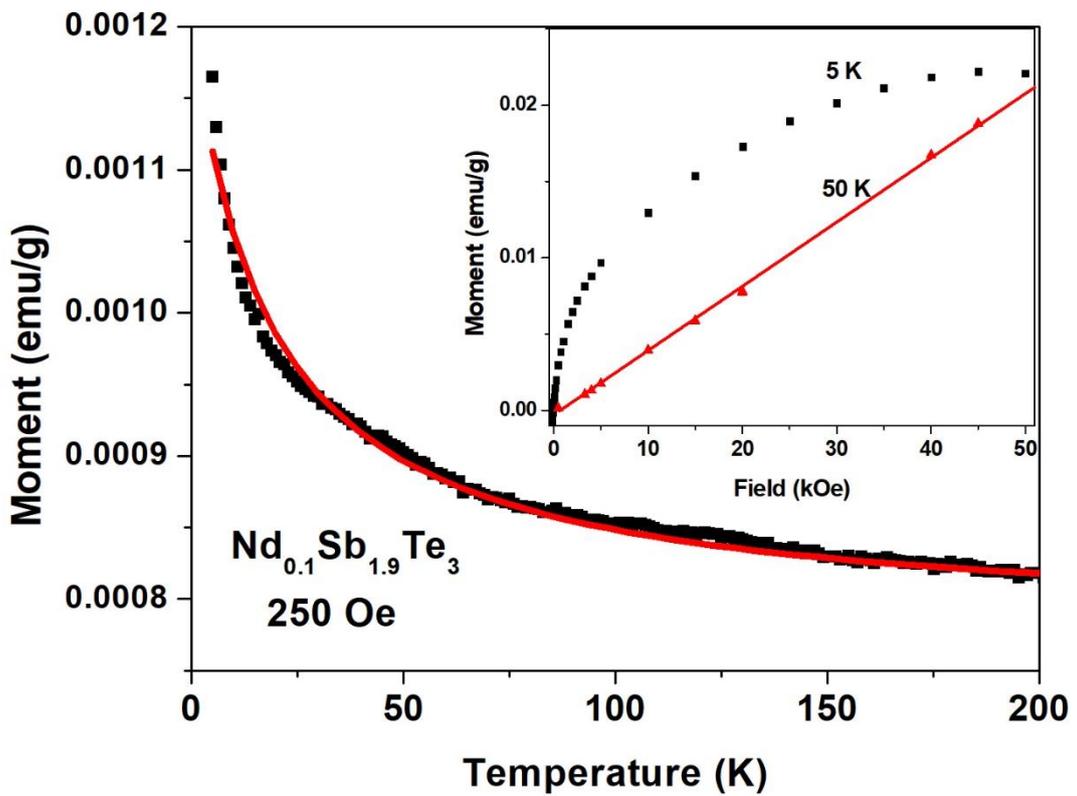

Fig 6

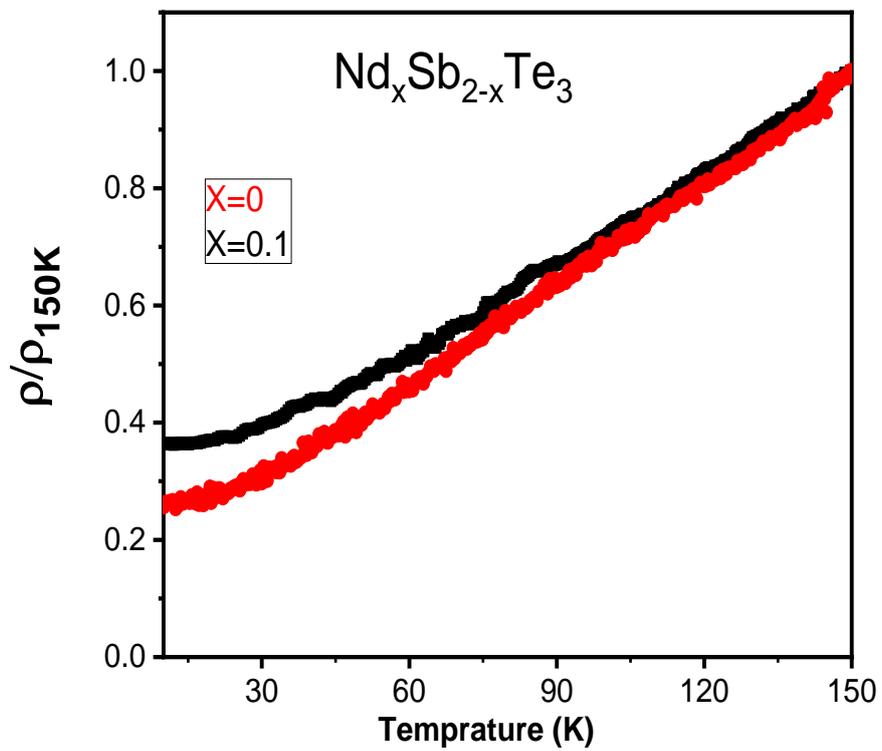

Fig 7(a)

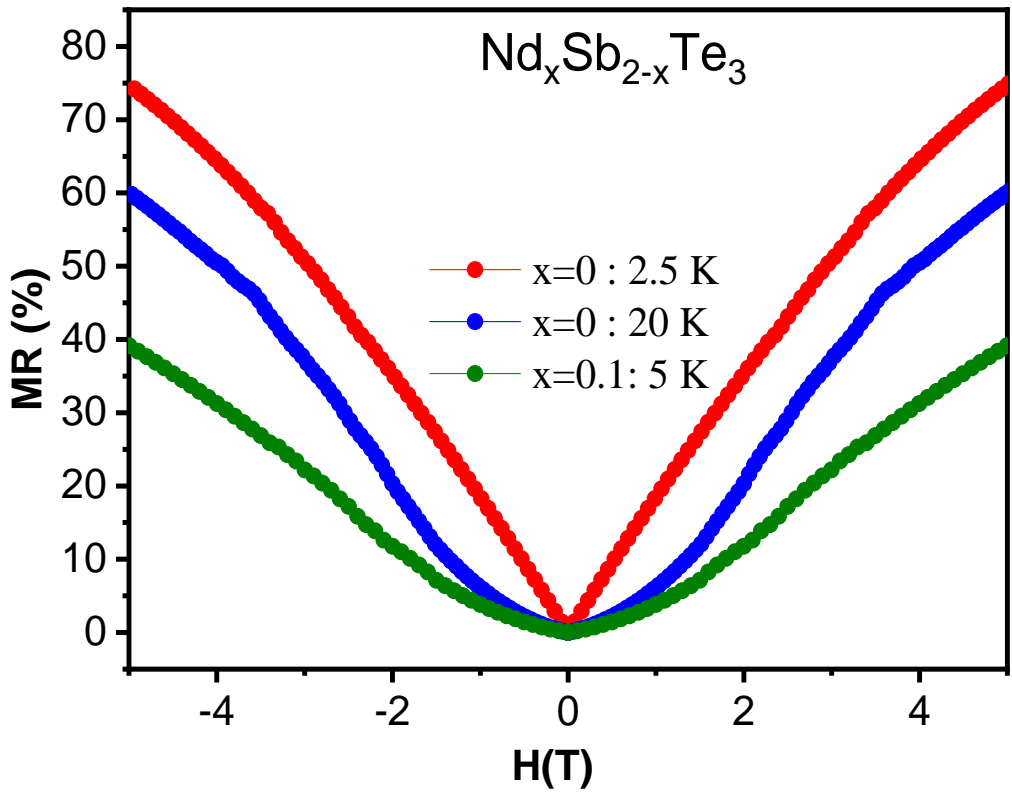

Fig 7(b)

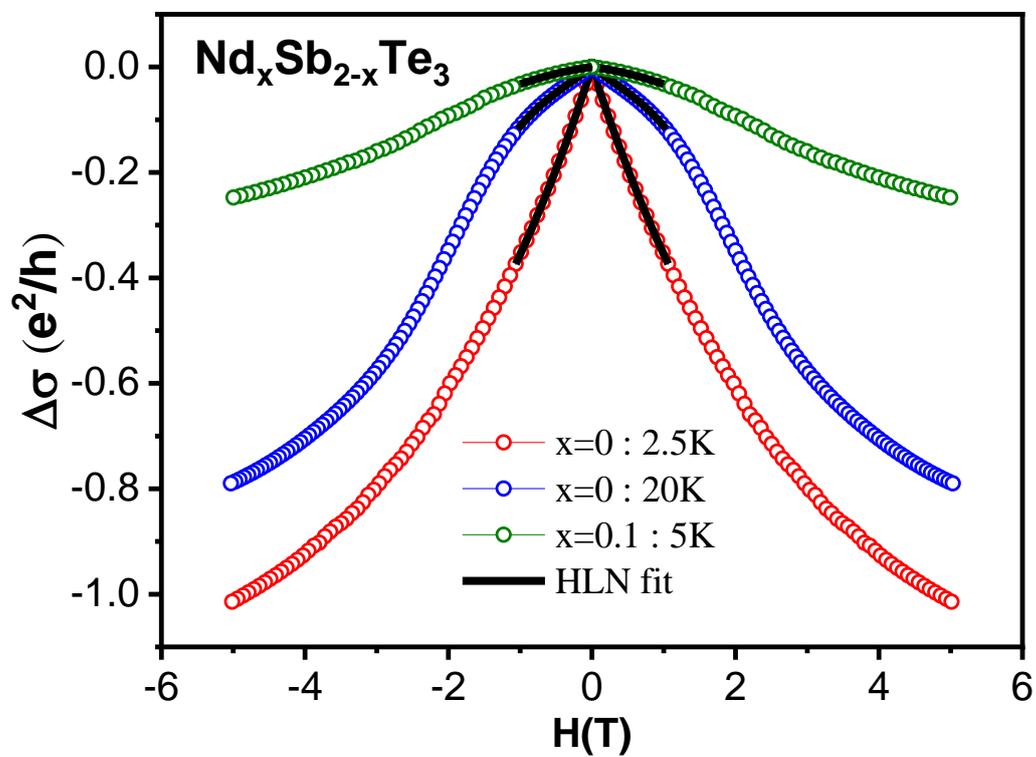